\documentclass[12pt]{article}
\usepackage{amsmath,amssymb}
 \usepackage{graphicx}

\usepackage{graphicx}
\usepackage{wrapfig}
\usepackage{color}

\usepackage{hyperref}
\usepackage[sort,compress]{cite}

\def\beq{\begin{eqnarray}}
\def\eeq{\end{eqnarray}}

\def\lb{\label}

\newcommand{\be}{\begin{equation}}
\newcommand{\ee}{\end{equation}}
\newcommand{\bea}{\begin{eqnarray}}
\newcommand{\eea}{\end{eqnarray}}
\newcommand{\bg}{\begin{gather}}

\newcommand{\bseq}{\begin{subequations}}
\newcommand{\eseq}{\end{subequations}}

\renewcommand{\ln}{\mathop{\rm ln}\nolimits}

\def\be{\begin{eqnarray}}
\def\ee{\end{eqnarray}}
\def\lb{\label}

\begin{document}

\title{\textbf{
Space-time structure, asymptotic radiation \\
and information recovery for \\ a quantum 
hybrid state}}
\vspace{0.5cm}
\author{ \textbf{ Yohan Potaux$^{1}$, Debajyoti Sarkar$^{2}$ and Sergey N. Solodukhin$^{1}$ }} 

\date{}
\maketitle
\begin{center}
    \emph{$^{1}$Institut Denis Poisson UMR 7013,
  Universit\'e de Tours,}\\
  \emph{Parc de Grandmont, 37200 Tours, France} \\
\vspace{0.4cm}
\emph{$^{2}$Department of Physics}\\
  \emph{Indian Institute of Technology Indore}\\
  \emph{Khandwa Road 453552 Indore, India}
\end{center}




\vspace{0.2mm}

\begin{abstract}
\noindent{ 
A hybrid quantum state is a combination of the Hartle-Hawking state for
the physical particles and the Boulware state for the non-physical ones (such as ghosts), as was introduced in  our earlier work \cite{Potaux:2021yan}. We present a two-dimensional example, based on the RST model,  when the corresponding back-reacted spacetime is a causal diamond, geodesically complete
and free of the curvature singularities. In the static case it shows no presence of the horizon while it has a wormhole structure mimicking the
black hole. In the dynamical case, perturbed by a pulse of classical matter, there appears an apparent horion while  the spacetime remains to be 
a regular causal diamond. We compute the asymptotic radiation both in the static and dynamic case. We define entropy of the asymptotic radiation
and demonstrate  that as a function of the retarded time it shows the behavior typical for the Page curve.
We suggest interpretation of our findings in terms of correlations in the virtual pairs of physical and non-physical particles spontaneously created in the spacetime.}
\end{abstract}


\rule{7.7 cm}{.5 pt}\\
\noindent ~~~ {\footnotesize   yohan.potaux@lmpt.univ-tours.fr,\\ dsarkar@iiti.ac.in,\\ Sergey.Solodukhin@lmpt.univ-tours.fr}

\pagebreak

\noindent{\bf 1. Introduction.}  In the last few decades the focus on the study of black holes has moved from their classical properties to their quantum aspects. One of the most discussed issues is the problem of the loss of the information and quantum unitarity.
In the present note, in an attempt to answer these important questions, we follow the same route that we started in  \cite{Potaux:2021yan}  and \cite{Berthiere:2017tms} some time ago. The key idea in this approach is that we first have to look, in a non-pertubative 
way, at the geometry that arises in the semi-classical description of the quantum gravity and that replaces the classical black hole geometry.
The semi-classical  gravitational equations that are supposed to be solved come from variation of the gravitational effective action
obtained after integrating out the quantum matter fields while keeping the metric classical. The respective effective action is very non-local
and in four dimensions we do not have a complete information about the full structure of this action although we do know how it can be represented as a formal series in curvature, at least to some lower order.  This in particular motivates the study of the two-dimensional gravity. In two dimensions the one-loop non-local effective action is well-known to be the famous Polyakov action. The non-locality present in the effective action
manifests itself in certain ambiguities when one tries to solve the respective gravitational equations. These ambiguities are 
usually understood as  some freedom in the choice of the quantum state.
Some of the quantum states that have been discussed in the literature are as follows.

\medskip

\noindent{\it The Hartle-Hawking state}:  it contains  thermal radiation at infinity and the stress-energy  tensor is regular at the horizon. It describes a black hole in thermal equilibrium with the Hawking radiation.
 
  \vspace{5mm}
  
\noindent {\it The Boulware state}: the stress-energy tensor is  vanishing  at infinity and there is no radiation there. However, being considered on a classical black hole metric, it is singular at the horizon.

\medskip

The Hartle-Hawking (HH) state is the most natural choice for the quantum particles with the physical spectrum that are in equilibrium with the
classical black hole. On the other hand, if non-physical particles are present, such as ghosts, the Boulware (B) state is the most appropriate one to describe these particles since it is the only one where one does not detect the flow of the non-physical particles at the asymptotic infinity of a classical black hole. This is similar to how the ghosts are incorporated in the 
quantum field theory: they are allowed to participate in the intermediate interactions but do not show up in the asymptotic states
or in the elements of the $S$-matrix. This is the proposal we made in our previous joint  paper \cite{Potaux:2021yan}. Notice that earlier the problem of ghosts was discussed in \cite{Strominger:1992zf}, see also \cite{Bilal:1992kv}.
In fact, if both physical and non-physical particles are present the adequate quantum state is a hybrid one:
the physical particles are in the Hartle-Hawking state while the non-physical ones are in the Boulware state.
The other observation, made on the case by case analysis in \cite{Potaux:2021yan},  is that if at least part of the particles
(either physical or non-physical) are in the Boulware state the respective back-reacted geometry is horizon-less.
A narrow throat, in which the $(tt)$ component of the metric is extremely small while non-vanishing, replaces the
classical horizon. The respective geometry then  becomes a black hole mimicker similar to those considered in the 4-dimensional case \cite{Damour:2007ap}.

Many aspects of black holes may be tested in the two-dimensional case \cite{Callan:1992rs} (for a review see \cite{Strominger:1994tn}).
In the present note we give a complete analysis, within the two-dimensional RST model \cite{RST}  (see \cite{RST-2} for a recent discussion), of the global structure of the back-reacted geometry and of the radiation at the asymptotic infinity in a particular case when the number of the non-physical fields exceeds the number of the physical fields.
In the parameter space of the semi-classical geometries analyzed in  \cite{Potaux:2021yan}   this case is special since it  corresponds to a completely regular spacetime with a global structure of the wormhole type. The spacetime is asymptotically flat at both ends of the wormhole.
Moreover, as we show below, the static spacetime is geodesically complete. This property persists in the dynamical case that
is modeled by sending in a pulse of the classical matter. In this case an apparent horizon is formed while the spacetime's global structure remains a causal diamond of the same type as the Minkowski spacetime.

The other particular goal of the present paper is to focus on the energy balance in the asymptotic regions. In the static case, even though
there is no horizon, the thermal radiation nevertheless is detected by an asymptotic observer. In the dynamical case
the radiation is thermal for earlier times but deviates from thermality as time passes. The deviations become
important when the observer approaches the apparent horizon due to  the arrival of the non-physical particles. 
The time variation in the radiation entropy that we define below then shows the typical behaviour for the Page curve.

\bigskip

\noindent{\bf 2. The classical model.}   
We choose the action describing a classical two-dimensional black hole to be a dilaton gravity of the form \cite{2d}, 
\begin{equation}
 {W}_0[g_{\mu\nu}, \phi] = \frac{1}{2\pi}\int \mathrm{d}^2x\, \sqrt{-g} \bigg\{e^{-2\phi}\big[R + 4(\nabla\phi)^2 + 4\lambda^2\big] - \frac{1}{2}(\nabla f)^2 \bigg\}
 \,.
 \lb{1}
\end{equation}
Here $\phi$ is the dilaton field which, together with the two-dimensional metric $g_{\mu\nu}$, form the gravitational degrees of freedom.
We also added the classical conformal matter described by the field $f(x)$ that will be useful to address the dynamical situation.
Variation of the action with respect to the metric is
\be
\pi T^{(0)}_{\mu\nu} 
 = 2e^{-2\phi}\Big\{-\nabla_\mu\nabla_\nu\phi + g_{\mu\nu}\big[\square\phi - (\nabla\phi)^2 + \lambda^2\big]\Big\}
 - \frac{1}{4}g_{\mu\nu}(\nabla f)^2
 + \frac{1}{2}\partial_\mu f \partial_\nu 
 = 0\,,
 \lb{1-1}
\ee
while the variation w.r.t. the dilaton is,
\be
 R + 4\big\{\square\phi - (\nabla\phi)^2 + \lambda^2\big\} = 0\,.
 \lb{1-2}
 \ee
In the absence of matter the solution of the theory is well known. We present it here in the light-cone coordinates $x^+$ and $x^-$,
\be
ds^2=-e^{2\rho}dx^+dx^-\, , \  \  e^{-2\rho}=e^{-2\phi}=\frac{M}{\lambda}-\lambda^2 x^+ x^-\, .
\lb{2}
\ee
A linear combination  of the equations (\ref{1-1}) and (\ref{1-2}) is $R+2\square\phi=0$, which explains why the conformal factor in (\ref{2}) is related to the dilaton field, $\rho=\phi$.
The metric is static since it admits a time-like Killing vector $\xi_\mu=\epsilon_\mu^{\ \nu}\partial_\nu \phi$. Its norm
$\xi^2= (\nabla\phi)^2 = -\frac{\lambda^4x^+x^-}{\frac{M}{\lambda} - \lambda^2x^+x^-}$ vanishes at $x^+=0$ and $x^-=0$.
This is the position of the Killing horizon. Notice that $(\nabla\phi)^2$ changes sign when one passes through the line $x^+=0$ (or $x^-=0$).
The scalar curvature of the metric $ R = \frac{4M\lambda^2}{M-\lambda^3x^+x^-}$ is divergent on the curve defined by equation $x^+ x^-=M/\lambda^3$. This is the position of the curvature singularity. The asymptotic infinity, where the dilaton $\phi\rightarrow -\infty$, lies for infinite values
$-x^+x^-\rightarrow +\infty$.
The parameter $M$ defines the mass of the black hole. The Hawking temperature
$T_H=\lambda/2\pi$ is set by the constant $\lambda$. This is an important difference from the four-dimensional case: the Hawking temperature is independent of the mass of the black hole. 

\bigskip

\noindent{\bf 3.  The hybrid RST model.}
In the quantum case the  classical action is modified by the terms that originate from integrating out the quantum conformal fields with the respective central charge $\kappa$. We here consider a case when there are two types of quantum fields with respectively positive and negative central charges, $\kappa_1>0$ and $\kappa_2<0$, that corresponds to the physical and non-physical fields.
The complete action is a version of the RST action
\be
&& W = W_0 + W_1 + W_2\,, \nonumber\\
 &&W_1 = \sum_{i=1}^2-\frac{\kappa_i}{2\pi}\int\mathrm{d} ^2x\,\sqrt{|g|}\bigg\{\frac{1}{2}(\nabla\psi_i)^2 + \psi_i R\bigg\} \,,\nonumber \\
 &&W_2 = - \frac{\kappa}{2\pi}\int\mathrm{d} ^2x\,\sqrt{|g|}\phi R\,, \  \, \kappa=\kappa_1+\kappa_2
\lb{3}
\ee
$W_1$ is a local form of the Polyakov action. We have introduced two different auxiliary fields $\psi_i\,,\,\,  i=1,2$. They both satisfy the same equation $\square \psi_i =R$ but with different boundary conditions.  The difference is due to the fact that the physical fields will be in the HH-state while the non-physical fields will be in the B-state. 
The variation of the complete action w.r.t. the metric is
\be
&&T^{(0)}_{\mu\nu}+T^{(1)}_{\mu\nu}+T^{(2)}_{\mu\nu}=0\,, \nonumber \\
 &&T^{(1)}_{\mu\nu} = \sum_{\kappa=1}^2\, \frac{\kappa_i}{\pi}\bigg\{\frac{1}{2}\partial_\mu\psi_i\partial_\nu\psi_i
 -\nabla_\mu\nabla_\nu\psi_i
 +g_{\mu\nu}\bigg(\square\psi_i - \frac{1}{4}(\nabla\psi_i)^2\bigg)\bigg\}
 \,, \nonumber \\
 &&T_{\mu\nu}^{(2)} = \frac{\kappa}{\pi}(g_{\mu\nu}\square\phi - \nabla_\mu\nabla_\nu\phi)
 \,. 
 \label{4}
 \ee
The equation for the dilaton becomes 
\be
 R\bigg(1 + \frac{\kappa}{2}e^{2\phi}\bigg) + 4\big\{\square\phi - (\nabla\phi)^2 + \lambda^2\big\} = 0\,.
 \lb{5}
 \ee
A linear combination of (\ref{5}) and of the trace of equation (\ref{4}) again leads to equation
 $R +2\square\phi=0$. This again allows us to use the gauge $\rho=\phi$.

\bigskip

\noindent{\bf 4. Conformal gauge and the solution.} 
Each field $\psi_i$ can be expressed as
\be
 \psi_i = -2\phi + w_i
 \,, \quad
 \square w_i=0
 \,, \quad w_i = w_{i+}(x^+) + w_{i-}(x_-)
 \,.
 \lb{6}
 \ee
and the components of $T^{(1)}_{\mu\nu}$ become
\be
 && T^{(1)}_{\pm\pm}  = \frac{2}{\pi}\sum_{i=1}^2\kappa_i\bigl\{\partial_\pm^2\phi - (\partial_\pm\phi)^2 - t_{i\pm}\bigr\}
  \,, \nonumber \\
  &&T^{(1)}_{+-}  = -\frac{2\kappa}{\pi}\partial_+\partial_-\phi
  \,,
 \lb{7}
\ee
where $t_{i\pm} \equiv \frac{1}{2}w_{i\pm}'' - \frac{1}{4}(w_{i\pm}')^2$.
We also define $T^{(12)}_{\mu\nu} \equiv T^{(1)}_{\mu\nu} + T^{(2)}_{\mu\nu}$ such that
\be
&&  T^{(12)}_{\pm\pm}  = \frac{1}{\pi}\sum_{i=1}^2\kappa_i\bigl\{\partial_\pm^2\phi - 2t_{i\pm}\bigr\}
  \,, \nonumber \\
 &&T^{(12)}_{+-}  = -\frac{\kappa}{\pi}\partial_+\partial_-\phi
  \,.
 \lb{8}
\ee
We define a new field
$ \Omega(\phi) \equiv  e^{-2\phi}+ \kappa \phi $.
The equations of motion give equations for $\Omega$,
\be
 && \partial_\pm^2\Omega -2\sum_{i=1}^2\kappa_i t_{i\pm} + \frac{1}{2}(\partial_\pm f)^2 = 0\,, \\
 && -\partial_+\partial_-\Omega - \lambda^2 = 0\, \nonumber
 \,.
 \lb{11}
\ee
The solution is found to be of the form
\be
 &&\Omega = -\lambda^2x^+x^- + u_+(x^+) + u_-(x^-)\,, \nonumber \\
 &&u_\pm'' = 2\sum_{i=1}^2\kappa_i t_{i\pm} - \frac{1}{2}(\partial_\pm f)^2\,. 
 \lb{12}
\ee
In the absence of  matter $(f=0)$ one finds  a static solution for which $u_+(x^+) + u_-(x^-) = \tilde{u}(x^+x^-)$. It  can be written in the form
\be
&& \Omega =
 -\lambda^2x^+x^- + 2(\kappa_1 P_1 + \kappa_2 P_2)\ln|\lambda^2x^+x^-| + \frac{M}{\lambda}
 \, , \nonumber \\
 &&u_\pm'' = -\frac{2(\kappa_1 P_1 + \kappa_2P_2)}{(x^\pm)^2} = 2(\kappa_1 t_{1\pm} + \kappa_2 t_{2\pm})\,.
 \lb{14}
\ee
Therefore we can take for a general pair $(\kappa_1, \kappa_2)$:
 $t_{i\pm} = - \frac{P_i}{(x^\pm)^2}$.
The choice of $t_{i\pm}$ corresponds to the choice of the quantum state. The appropriate analysis was done in \cite{Potaux:2021yan}.
The physical fields (with $\kappa_1>0$) are in the Hartle-Hawking state (in the present context discussed in \cite{Solodukhin:1995te}),
which means no singularity at the classical horizon ($x^+x^-=0$). This fixes $P_1=0$. On the other hand, the non-physical fields (with $\kappa_2<0$)
are in the Boulware state. No radiation of the non-physical fields at asymptotic infinity ($-x^+x^-\rightarrow +\infty$) implies that $P_2=-\frac{1}{4}$. 
Thus, in the static case, the dilaton $\phi(x^+x^-)$ is a solution of the master equation,
\be
 \Omega(\phi) = -\lambda^2x^+x^- - \frac{\kappa_2}{2}\ln|\lambda^2x^+x^-| + \frac{M}{\lambda}
 \,,
 \lb{16}
\ee
where $\Omega(\phi)= e^{-2\phi} + \kappa\phi$ (we remind that $\kappa<0$).

\bigskip

\noindent{\bf 5. The global structure of the static solution for $\kappa<0$.}
Although the values of $\kappa_1>0$ and $\kappa_2<0$ can be arbitrary, the most interesting case is when the total
$\kappa=\kappa_1+\kappa_2$ is negative. It corresponds to a totally regular spacetime, as we show below. The case of general $\kappa$ will be considered in detail in an upcoming paper by the same authors.

We first consider the causal diamond defined as ${\cal M}\equiv (x^+\geq 0, x^- \leq 0)$. In this domain $(-\lambda^2 x^+ x^-)\geq 0$ and the function that stands in the right hand side of the master equation (\ref{16}) is monotonically growing from $-\infty$ to $+\infty$ as 
$(-\lambda^2 x^+ x^-)$ changes from $0$ to $+\infty$. Similarly, the function $\Omega(\phi)= e^{-2\phi}+\kappa \phi$ is monotonically growing as
$\phi$ changes from $+\infty$ to $-\infty$ if $\kappa=\kappa_1+\kappa_2$ is negative. Thus for any value of $x^+ x^-$ in this domain there exists
only one value of the dilaton $\phi$, i.e. it is a single valued function here. 

Two limits are of particular interest.

\medskip

1). $\phi\rightarrow -\infty$ (or $-x^+ x^-\rightarrow +\infty)$:
\be
e^{-2\phi}=-\lambda^2 x^+ x^-+\frac{\kappa_1}{2}\ln(-\lambda^2 x^+ x^-)+M/\lambda +\dots\,.
\lb{17}
\ee

\medskip

2).  $\phi\rightarrow +\infty$ (or $-x^+ x^-\rightarrow 0)$:
\be
\phi=-\frac{\kappa_2}{2\kappa}\ln (-\lambda^2 x^+ x^-)-(-\lambda^2 x^+ x^-)^\frac{\kappa_2}{\kappa}e^{\frac{2M}{\lambda \kappa}}\dots\,,
 \lb{18}
 \ee
 where $\dots$ stand for the subleading terms.
 
 \bigskip
 
 \noindent{\it 5.1 Asymptotic flatness.} In both these limits the spacetime is asymptotically flat. This can be shown by directly computing the scalar curvature, using the general formula,
$
R=8e^{-2\phi}\partial_+\partial_-\phi\,.
$
One finds,
\be
R=\left\{
 \begin{aligned}
  O\left(\frac{\kappa_1}{x^+ x^-}\ln(-\lambda^2 x^+x^-)\right)\, , \ \  -x^+ x^-\rightarrow +\infty
  \,, \\
  O\left((-x^+ x^-)^\frac{\kappa_2-\kappa_1}{\kappa}\right)\, , \ \ \  \ \ \ \ \     -x^+ x^-\rightarrow 0
  \,,
 \end{aligned}
 \right.
\lb{20}
\ee
where we use that ${(\kappa_2-\kappa_1)}/{\kappa}>0$.

\bigskip

\noindent{\it 5.2 Geodesic completeness.} For a geodesic that at fixed value of $x^+_0$ goes from some initial value $x^-_0$ to value $x^-$ the change in the affine parameter $\chi$ is given by equation
\be
\Delta \chi=A \int_{x_0^-}^{x^-}du\, e^{2\phi(u, x_0^+)}\,,
\lb{21}
\ee
where $A$ is an integration constant. A similar equation is valid for a geodesic at fixed value of $x^-_0$ that goes along the $x^+$-direction.
The variation in the affine parameter is obviously divergent in this direction when the end point approaches $x^+=+\infty$.
This is the classical region where $e^{2\phi}\sim 1/(-\lambda^2 x^+ x^-)$ and the spacetime is expected to be geodesically complete there.
For a geodesic that goes to the other side of the future null infinity in the causal diamond $\cal M$, $x^-\rightarrow 0$, the integral (\ref{21}) is also divergent since $e^{2\phi} = O((-\lambda^2 x^+ x^-)^{-\kappa_2/\kappa})$.
We can also check this by performing a change of coordinates. Let us define
\begin{equation}
 y^\pm = \pm\lambda^{-\kappa_2/\kappa}e^{M/\lambda\kappa}\frac{\kappa}{\kappa_1}\frac{1}{(\pm x^\pm)^{-\kappa_1/\kappa}}
 \,,
 \lb{22}
\end{equation}
such that $ d s^2 \sim -d y^+d y^-$
when $x^\pm \rightarrow 0^\pm$, so the metric is regular in this limit which corresponds to $y^\pm \rightarrow \mp \infty$.
In these coordinates the spacetime metric is given by
\begin{equation}
 d s^2 = -e^{-2M/\lambda\kappa}e^{-2\phi}(-\lambda^2x^+x^-)^{\kappa_2/\kappa} dy^+ dy^-
 \,.
\end{equation}

We conclude that the causal diamond $\cal M$ is geodesically complete and represents the entire spacetime.
However there could be other geodesically complete domains such as $\tilde{\cal M}\equiv (x^+>0, x^->0)$.

\bigskip

\noindent{\it 5.3 No Killing or apparent horizon.} The static space-time metric admits a time-like Killing vector
$\xi=\lambda (x^+\partial_+-x^-\partial_-)$. Its norm $\xi^2=\lambda^2x^+x^-e^{2\phi}$ is everywhere negative. 
It may potentially vanish at $x^+ x^- =0$. However, the analysis, using the asymptotic behaviour of the dilaton (\ref{18}), shows that 
$-\xi^2 = O((-\lambda x^+ x^-)^{1-\frac{\kappa_2}{\kappa}})$ diverges  as one approaches $x_+ x_-=0$ due to $1-\frac{\kappa_2}{\kappa}<0$.

There may also exist an apparent horizon that can be defined by the condition that the norm of the gradient of the dilaton vanishes.
The calculation shows that 
\be
(\nabla\phi)^2=-\frac{4 e^{-2\phi}}{(2 e^{-2\phi}-\kappa)^2}\frac{(-\lambda^2x^+ x^- -\kappa_2/2)^2}{x^+ x^-}\,.
\lb{21}
\ee
It vanishes when $-\lambda^2x^+ x^-=\kappa_2/2<0$. However, this is outside the domain ${\cal M}\equiv(x^+\geq 0, x^-\leq 0)$.
The gradient may vanish when $\phi$ approaches $+\infty$ (or $x^+x^-\rightarrow 0$). It is indeed the case,
$(\nabla\phi)^2\sim (-x^+ x^-)^{(\frac{\kappa_2}{\kappa}-1)}\rightarrow 0$ since $\kappa_2/\kappa>1$. However, this happens on the border of the
domain $\cal M$ and hence it can not define a boundary of the trapped region.

\begin{figure}
\begin{center}
  \includegraphics[width=110mm]{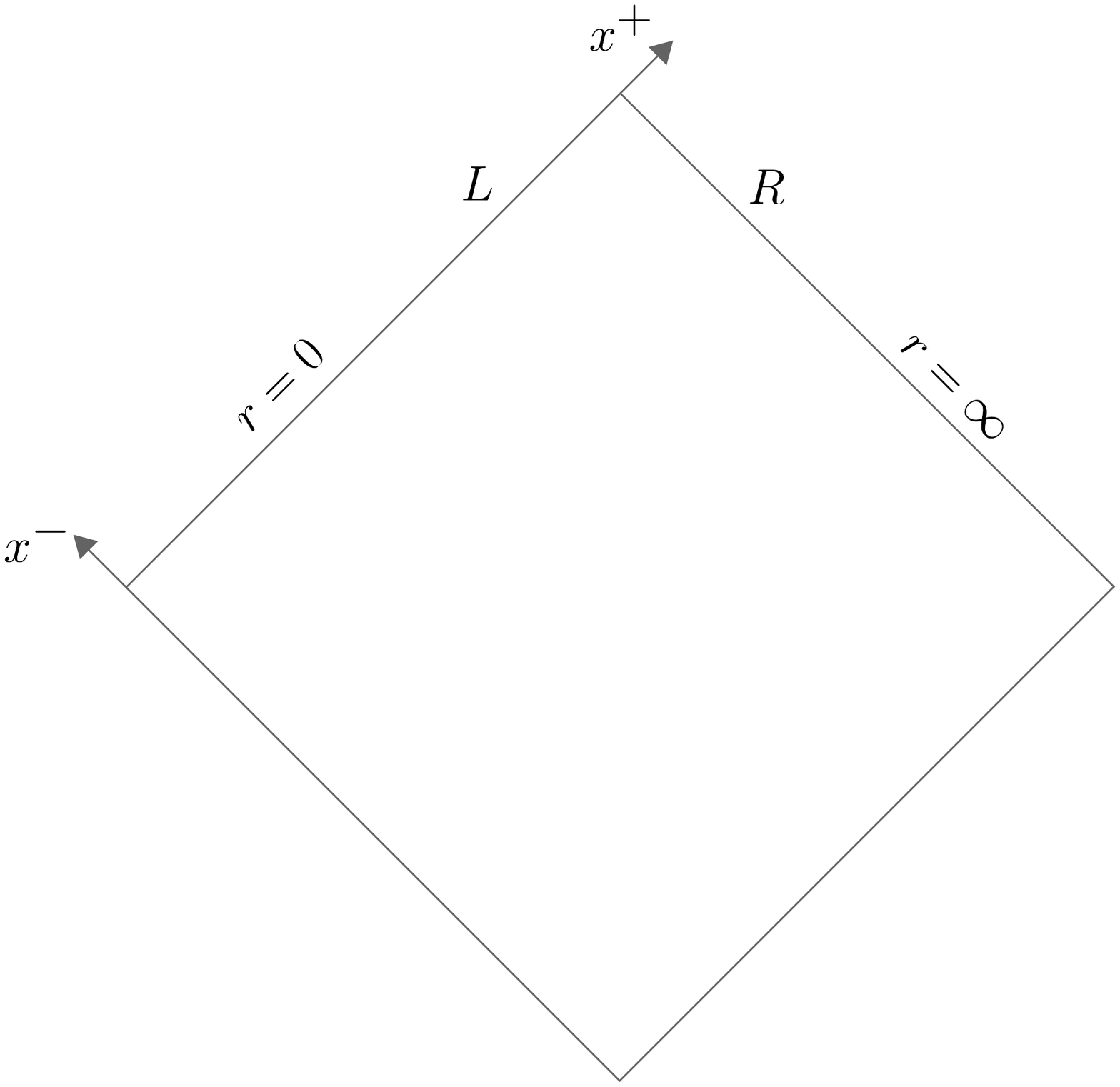}
  \end{center}
  \caption{\rm Causal diamond structure of the static solution. Left (L) and right (R) future infinity with limiting value of $r=e^{-\phi}$.}
  \label{Fig1}
\end{figure}

Thus we conclude that in the domain $\cal M$ there exists neither Killing nor apparent horizons. The spacetime $\cal M$ is geodesically 
complete and horizon-less. It has the global structure of a causal diamond similar to the Minkowski spacetime (see figure \ref{Fig1}).

\bigskip

\noindent{\it 5.4 Wormhole structure (black hole mimicker).}
Let us define the $(tt)$ metric function $g(x^+x^-)=-\xi^2$ as minus the norm of the Killing vector $\xi$. As we have discussed  above it never vanishes and hence is positive everywhere in $\cal M$. 
It approaches infinity when $-x^+x^-\rightarrow 0$ and approaches $+1$ from below when $-x^+x^- \rightarrow + \infty$. Hence it has to have a minimum somewhere in between. This minimum is found by imposing the condition $\partial_+ g=0$
(or, equivalently, $\partial_- g=0$). Since $g=-\lambda^2 x^+ x^- e^{2\phi}$ this condition reduces to condition on the dilaton,
$x^+\partial_+ \phi=-\frac{1}{2}$.  Differentiating the master equation (\ref{16}) and imposing the above condition one finds that at the minimum the dilaton takes the value $e^{-2\phi_{\rm min}}=\frac{\kappa_1}{2}-\lambda x^+ x^-$ where the value of $x^+ x^-$ at the minimum is found by solving equation
\be
\left(\frac{\kappa_1}{2}-\lambda^2 x^+ x^-\right)^{-\kappa}=(-\lambda^2 x^+ x^-)^{-\kappa_2} e^{2a-\kappa_1}\,,
\lb{22}
\ee
where we introduced $a=M/\lambda$.  This equation can be solved perturbatively assuming that at the minimum $-\lambda^2 x^+ x^-\ll \frac{\kappa_1}{2}$. 
This will be the case if $a\gg a_{cr}=-\frac{\kappa_1}{2}\ln \frac{\kappa_1}{2e}$.
Then one finds that at the minimum $e^{-2\phi_{\rm min}}\simeq \frac{\kappa_1}{2}$ and $-\lambda^2 x^+ x^-\simeq (\frac{\kappa_1}{2})^{\kappa/\kappa_2}e^{\frac{2a-\kappa_1}{\kappa_2}}$
and the $(tt)$ metric function is
\be
{\rm min}_{\cal M}\, g\simeq \bigg(\frac{2e}{\kappa_1}\bigg)^{-\frac{\kappa_1}{\kappa_2}}e^{-S_{cl}/|\kappa_2|}
\,,
\lb{23}
\ee
where $S_{cl}=2a$ is the entropy of the classical black hole. Thus, we see that the spacetime $\cal M$ has the structure of a wormhole, the region where $g$ takes its minimum is defined as the throat in which the time flows extremely slowly with respect to the time of an external observer.
This is a direct realisation of a black hole mimicker  as suggested in \cite{Damour:2007ap}. The throat appears as a replacement for the classical black hole horizon when one imposes the Boulware state condition for a part of the quantum fields. Notice that at the minimum (\ref{23}) the metric function  is exponentially small in terms of the classical black hole entropy, as was predicted in \cite{BTZ} (see also \cite{Germani:2015tda}).

\bigskip

\noindent{\bf 6. The global structure of a dynamic solution for $\kappa<0$.}
Let us consider the same domain $\{x^+>0, x^-<0\}$ with a shock wave of classical  matter traveling along the geodesic $x^+=x^+_0$
in the $x^-$ direction with mass $m>0$,
 $\frac{1}{2}(\partial_+f)^2 = \frac{m}{\lambda x^+_0} \delta(x^+-x^+_0)$.
Then, integrating equation (\ref{12}),  we find for the master equation, for $x^+\geq x_0^+$,
\begin{equation}
 \Omega(\phi)  = -\lambda^2x^+\left(x^-+ \frac{m}{\lambda^3x_0^+}\right)- \frac{\kappa_2}{2}\ln(-\lambda^2x^+x^-) + \frac{M+m}{\lambda} 
 \,.
 \lb{24}
\end{equation}
The spacetime above the line $x^+=x^+_0$ is no more static. It becomes evolving and an apparent horizon appears.

\bigskip

\begin{figure}
\begin{center}
  \includegraphics[width=110mm]{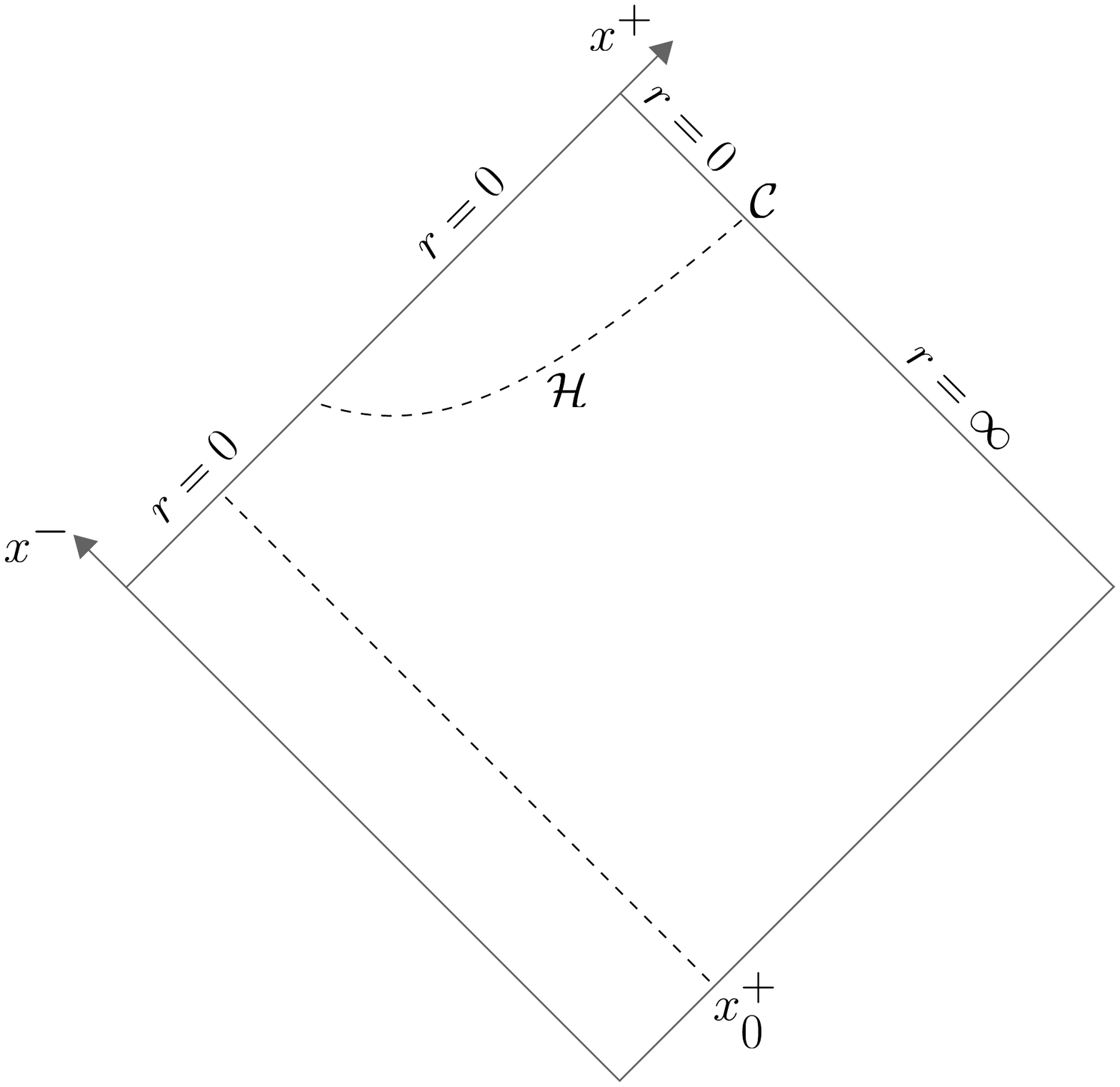}
  \end{center}
  \caption{\rm Causal diamond structure of the dynamic solution. Apparent horizon $\cal H$ intersecting the right future infinity at $\cal C$.}
  \label{Fig2}
\end{figure}

\noindent{\it 6.1 Apparent horizon.} The equation $(\nabla\phi)^2=0$ now has a solution in the domain $\{x^+ > 0, x^- < 0\}$. Indeed the condition $\partial_-\phi=0$ leads to the equation $-\lambda^2x^+ x^-=\frac{\kappa_2}{2}<0$ which describes a curve that lies 
outside of the domain  $\{x^+>0, x^-<0\}$, as in the static case. However, the condition 
$\partial_+\phi=0$ defines a curve
\be
{\cal H}: \  \  -\lambda^2x^+\left(x^-+\frac{m}{\lambda^3x_0^+}\right)=\frac{\kappa_2}{2}<0\, ,  \ \ \ x^+\leq x^+_0\, 
\lb{25}
\ee
which has a branch in the domain  $\{x^+>0, x^-<0\}$.  This curve starts at the point  $\{ x^-=0, x^+=x^+_h={(-\kappa_2)\lambda x_0^+}/{2m}\}$ and ends at the point $\{x^-=x^-_h=-\frac{m}{\lambda^3x_0^+}, x^+=+\infty\}$, see figure \ref{Fig2}. For small $m$ the region inside the curve $\cal H$ represents just a small portion of the spacetime ${\cal M}$ near the conner $\{ x^-=0, x^+=+\infty\}$.

\bigskip

\noindent{\it 6.2 Asymptotic behaviour of the dilaton and the scalar curvature.} 
There are several asymptotic regions where we are interested to know the new asymptotic behavior of the dilaton and the curvature.

\medskip

i). $(x^-=0, \,\text{finite}\,\,x^+)$.  In the limit $x^-\rightarrow 0$, keeping the value of $x^+$ finite, one finds from the master equation (\ref{24}) that the dilaton $\phi$ goes to $+\infty$ irrespective of the value of $x^+$. The asymptotic behaviour of the dilaton there is still given by
equation (\ref{18}) where $M$ has to be replaced by $M+m(1-x^+/x^+_0)$. The scalar curvature vanishes in this limit for any finite value of
$x^+$. Thus, the dynamic spacetime is still asymptotically flat in the left future null infinity.

\medskip

ii). $(x^-< x^-_h,\,\, x^+=+\infty)$. This is the part in the asymptotic  region that is outside the apparent horizon $\cal H$. Here the dilaton goes to $-\infty$, just as in the static case. Its asymptotic behavior is similar to the one given in (\ref{17}), 
\be
e^{-2\phi}=-\lambda^2 x^+ (x^--x_h^-)+\frac{\kappa_1}{2}\ln(-\lambda^2 x^+ x^-)+\dots\,.
\lb{26}
\ee
Scalar curvature vanishes in this asymptotic regime, $R\rightarrow 0$ just as in the static case.

\medskip

iii). $(x^-_h<x^-<0,\,\, x^+=+\infty)$.  This is the asymptotic region that lies inside the apparent horizon $\cal H$. Asymptotically, the dilaton 
goes to $+\infty$. The asymptotic behavior  of the dilaton and the scalar curvature is 
\be
&&\phi=\frac{\lambda^2}{\kappa} x^+(x^--x^-_h)+\dots\,, \\
&&R=-\frac{8\lambda^2}{\kappa}e^{2\lambda^2x^+(x^--x^-_h)/\kappa}+\dots\,.\nonumber
\lb{27}
\ee
So that the spacetime is  asymptotically flat inside the apparent horizon $\cal H$ at both sides. 

\medskip

iv). $(x^-=x^-_h,\,\, x^+=+\infty)$. This is a special point  $\cal C$ where the apparent horizon $\cal H$ crosses the future null infinity. The asymptotic behavior of the dilaton in this case is 
\be
e^{-2\phi}=-\frac{\kappa_2}{2}\ln(-\lambda^2x^+ x^-)+\dots\,,
\lb{28}
\ee
so that $\phi$  approaches $-\infty$ there.

 To analyse the curvature in this limit, we first compute all derivatives of the dilaton and only then take the limit. It turns out that the curvature approaches a constant value $R=4\lambda^2$. Thus, the curvature is not smooth
on the future null infinity: it vanishes everywhere  except for a single point $\cal C$.

\bigskip

\noindent{\it 6.3 Geodesic completeness.}   The analysis shows that  all null geodesics can be continued indefinitely, 
the variation in the affine parameter is infinite approaching the future null infinity, both outside and inside the apparent horizon.
A special attention is required for a null geodesic that goes along the $x^+$ direction at $x^-=x^-_h$, approaching the point $\cal C$.
A variation in the affine parameter for this geodesic is obtained, using equation (\ref{28}), by integrating 
\be
\int^{x^+}dv \,e^{2\phi(x^-_h, v)}\simeq \frac{2 x^+}{-\kappa_2 \ln(-\lambda^2 x^+ x^-_h)}\rightarrow +\infty \, , \  \  x^+\rightarrow \infty\,,
\lb{29}
\ee
where we used the asymptotic expansion for the integral logarithm $\int_0^x\frac{dx}{\ln x}={\rm li}(x)=x/\ln(x)+\dots$ valid for large $x$. Thus, the dynamical version of the spacetime $\cal M$ is geodesically complete.

\bigskip

\noindent{\bf 7. Asymptotic energy density.}
We consider the stress-energy $T^{(12)}_{\mu\nu}$ as the part due to the quantum matter while $T^{(0)}_{\mu\nu}$ is purely geometrical, similar to the Einstein tensor in the 4d gravitational equations. In the hybrid quantum state and considered on the solution to the
complete gravitational equations (\ref{4}), this part reads 
\be
2\pi T^{(12)}_{\pm\pm}=2\kappa\partial_\pm^2\phi-\frac{\kappa_2}{(x^\pm)^2}\, .
\lb{30}
\ee
Clearly, it has two contributions: one ($T^{(12,p)}_{\mu\nu}$) due to the physical fields proportional to $\kappa_1$ and the other ($T^{(12,np)}_{\mu\nu}$) due to the non-physical fields proportional to $\kappa_2$.
Some particular limits are of interest both in the static and the dynamic cases.

\bigskip
\noindent{\it 7.1 Static case.} 

\noindent  i) Left future null infinity ($x^-\rightarrow 0$ for a fixed $x^+$): 
\be
T^{(12)}_{++}=T^{(12, p)}_{++}+T^{(12,np)}_{++}=0\, ,
\lb{31}
\ee
so that in this asymptotic region the out-going radiation  of physical particles ($T^{(12, p)}_{++}>0$) and non-physical particles  ($T^{(12, np)}_{++}<0$)
exactly compensate each other. Note that the other component $T^{(12)}_{--}\sim (-x^-)^{\frac{\kappa_2}{\kappa}-2}$ does not necessarily vanish, when $x^-\rightarrow 0$. 

\medskip

\noindent  ii) Right future null infinity ($x^+\rightarrow +\infty$ for a fixed $x^-$):
\be
T^{(12)}_{++}=0\, , \ \ 2\pi T^{(12)}_{--}=\frac{\kappa_1}{(x^-)^2}\,.
\lb{32}
\ee
Thus only the physical particles contribute to the energy density, which is positive. This is expected as the hybrid quantum state is designed to have precisely this property. This is the energy density of the thermal gas with temperature $T=\lambda/2\pi$ as can be seen by passing to the asymptotic coordinates $(\sigma^+,\sigma^-): \ \lambda x^+=e^{\lambda \sigma^+}\, , \ \lambda x^-=-e^{-\lambda \sigma^-}$ in which metric is constant $ds^2=-d\sigma^+ d\sigma^-$. Thus, despite the absence of a horizon an outside observer detects  the thermal radiation.

\bigskip
\noindent{\it 7.2 Dynamic case.}  Compared to the static case the spacetime now is divided into two regions, inside and outside the apparent horizon.
As a result, there are additional asymptotic limits.

\medskip

\noindent i) Left future null infinity, both inside and outside apparent horizon: the stress energy vanishes in the same way as in  the static case, see (\ref{31}).

\medskip

\noindent  ii) Right future null infinity, inside apparent horizon  ($x^->x^-_h$): the stress energy vanishes, 
\be
T^{(12)}_{++}=0\, , \ \  T^{(12)}_{--}=0\, .
\lb{33}
\ee

\medskip
\noindent  iii) Right future null infinity, outside the apparent horizon ($x^-<x^-_h$):
\be
T^{(12)}_{++}=0\, , \ \  2\pi T^{(12)}_{--}=\frac{\kappa}{(\tilde{x}^-)^2}-\frac{\kappa_2}{(x^-)^2}\, ,
\lb{34}
\ee
where $\tilde{x}^-=x^--x^-_h$.
For large negative values of $x^-$ it approaches the positive thermal energy density (\ref{32}) while as $x^-$ increases there appears  a negative
contribution due to the non-physical particles. The above equation is derived under assumption that $x^+\rightarrow \infty$ while keeping 
$\tilde{x}^-$ fixed and non-zero. So that it can not be applicable close to the point $\cal C$ where $\tilde{x}^-=0$. The latter case has to be considered separately.

\medskip
\noindent iv) At the point $\cal C$  $(x^-=x^-_h\, , \ x^+\rightarrow \infty)$ on the apparent horizon: 
\be
T^{(12)}_{++}=0\, , \ \ 2\pi T^{(12)}_{--} \sim \kappa\,\left(\frac{\lambda^2 x^+}{\kappa_2\ln x^+}\right)^2\, .
\lb{35}
\ee
One sees that the energy density is divergent when approaching the point $\cal C$ on the future null infinity.

Effectively, inside the apparent horizon 
no out-going radiation appears in any asymptotic region there.
\bigskip

\noindent{\bf 8. Entropy of radiation and the Page  curve.} 
Entropy of the radiation in the right asymptotic region can be defined by the differential equation,
\be
\partial_-S=2\pi(-x^-)T^{(12)}_{--}\, .
\lb{36}
\ee
This equation is a differential form of the first law, $dS=T^{-1} dE$,  where $T=\lambda/2\pi$ and $dE=T_{--} d\sigma^-$ is the energy defined in terms of the coordinates $(\sigma^+, \sigma^-)$ introduced above. 
Using equation (\ref{34}) one finds that for large negative values of $x^-$ the entropy $S$ is increasing. It is due to the positive energy density of the physical particles. Then there exists a point $x^-_m$ at which the entropy has maximum (its derivative $\partial_- S=0$).
This happens at $x^-_m=\left(1-\sqrt{\frac{\kappa}{\kappa_2}}\right)^{-1}x^-_h$ so that $x^-_m<x^-_h$. 
Then, for values $x^-> x^-_m$ the entropy is monotonically decreasing. In order to avoid the singularity at point $\cal C$
we may consider the change in the entropy along the line of large but finite value of $x^+$. Then $S(x^-)$ shows the behavior typical for the
Page curve. The total change in the entropy  along the line of constant $x^+$ while $x^-$ is changing in the interval $(-\infty, 0)$ is given by integral,
\be
\Delta S=2\pi \int_{-L}^{-\epsilon} dx^- (-x^-)T^{(12)}_{--}\, , \  \ \ x^+={\rm const}
\lb{37}
\ee
where we introduced two regulators, $L$ and $\epsilon$ to regularise the possible divergences for, respectively, large negative $x^-$ and $x^-$ close to $0$, and $T^{(12)}_{--}$ is given by (\ref{30}). The integration can be easily performed by a couple of integrations by parts and reduces to two boundary terms, at $x^-=-\epsilon$ and at $x^-=-L$. The computation shows that (\ref{37}) does not depend on $\epsilon$, provided it is small, so that one can safely take the limit $\epsilon\rightarrow 0$. On the other hand, the large $L$ regulator should be kept till the end. The result is
\be
\Delta S=\kappa_1\ln (\lambda^2 x^+ L)-\kappa_1+\frac{2M}{\lambda}+\frac{2m}{\lambda}\left(1-\frac{x^+}{x^+_0}\right)\,.
\lb{38}
\ee
\begin{figure}
\begin{center}
  \includegraphics[width=110mm]{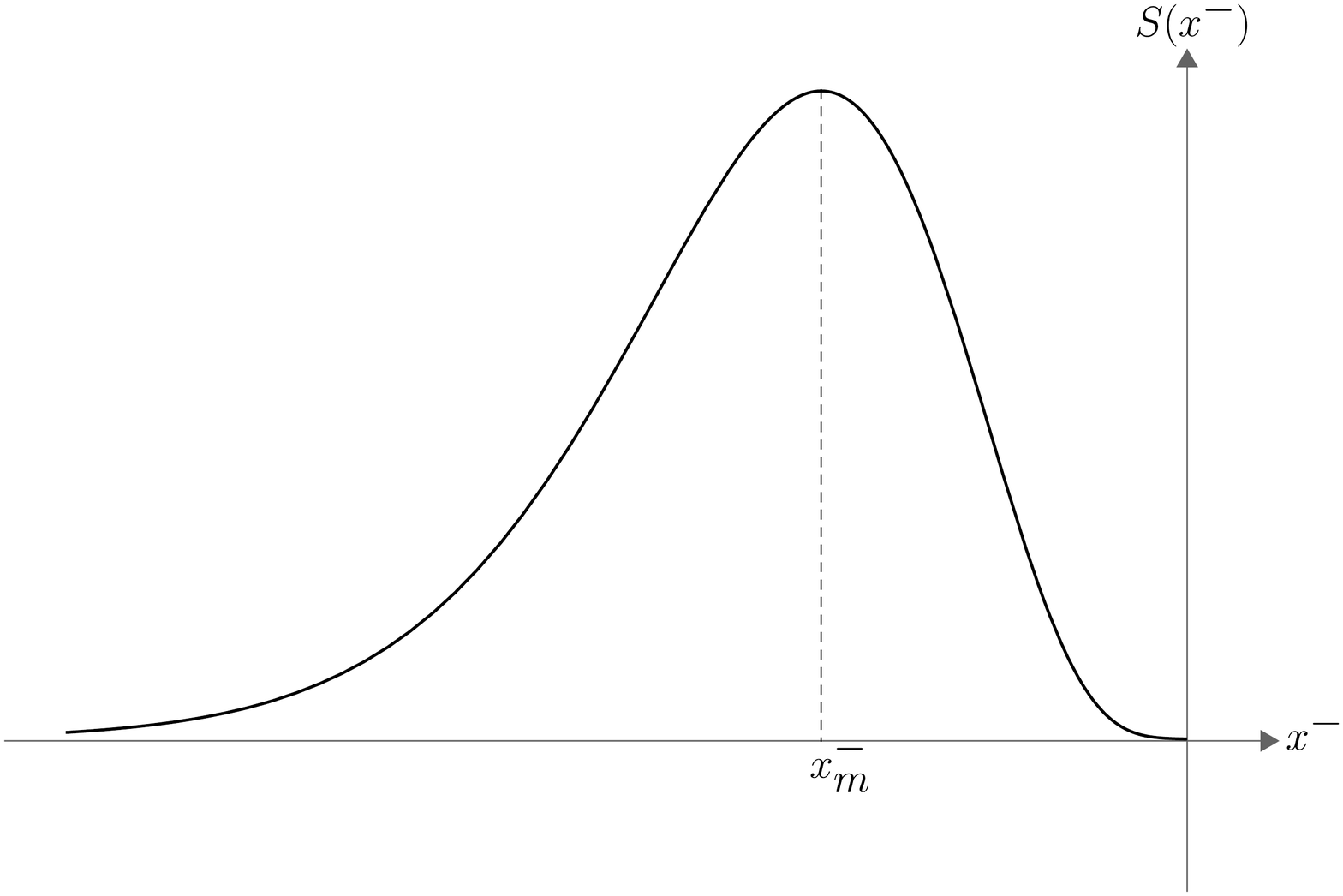}
  \end{center}
  \caption{\rm Asymptotic radiation entropy $S(x^-)$ as function of the retarded time $x^-$.}
  \label{Fig3}
\end{figure}
Notice that, rather surprisingly,  it contains the two last terms that are completely classical and do not depend on $\kappa_1$.
Computationally, they come from the upper limit in the integral (\ref{37}).
The other observation is that (\ref{38}) does not depend on $\kappa_2$ so that effectively only physical particles contribute to the total  change in the entropy. The total change (\ref{38}) vanishes for $x^+_L\simeq \frac{\lambda \kappa_1 x_0^+}{2m}\ln L$. In the limit $L$ to infinity
the integration curve can be adjusted at $x^+=x^+_L\rightarrow \infty$ so that the total change in the entropy is precisely zero.
This is precisely the behavior expected for the Page curve \cite{Page:1993wv} (for a recent discussion see \cite{Almheiri:2020cfm}). For an illustration of our result, see figure \ref{Fig3}.

\bigskip

\noindent{\bf 9. Interpretation and conclusions.} Here we give some qualitative (and inevitably non-rigorous)  interpretation of our results. They could be properly
understood in terms of creation of pairs of left- and right-moving pairs of physical and non-physical particles. This interpretation is in parallel to the well known and rather standard interpretation of the Hawking radiation in terms of pairs of particles of positive and negative energy that are created at the horizon so that the negative energy particle goes deep inside the hole while the positive energy particle escapes to infinity and forms the Hawking radiation. In our case the role of the negative energy particles is played by the non-physical particles. In the static case, when no horizon is actually present,  the spontaneously created left-moving (physical and non-physical) particles  reach the left future null infinity and precisely compensate  the  energy of each other so that the total energy flux vanishes, as in (\ref{31}). A similar pair of the right-moving particles behaves differently. The physical particles go to the right future null infinity and form the asymptotic radiation there,  which is given in (\ref{32}).
On the other hand, their non-physical counter-part stay in the bulk and do not reach the asymptotic region. They are in the Boulware quantum state
that guarantees that the non-physical particles are not visible in the asymptotic region in the static case.

In the dynamic case there appears an apparent horizon that changes the behavior of the right-moving  non-physical particles.
They now can reach the right asymptotic infinity but at much later retarded time $x^-$ than when they were spontaneously created.
So that there appears a delay between the physical particles coming out early (at sufficiently large negative values of $x^-$) and the
non-physical particles the majority of which are coming out much later, for $x^->x^-_m$. For the entropy of the radiation this explains the declining part of the curve $S(x^-)$. 

Speaking in terms of  information, at the moment of spontaneous creation, the physical and non-physical particles in the  pairs  are correlated to
each other. These correlations, for earlier retarded time $x^-$ appear to be  lost for a distant observer since only physical particles are visible at infinity. They can be reconstructed for later times   when the non-physical particles start to arrive in the asymptotic region. So that when all particles come out at infinity the information gets totally reconstructed.   

Two lessons could be learned from the two-dimensional example considered here. First, when the back reaction is taken into account, the presence of particles in the Boulware quantum state may drastically change the global geometry of the spacetime, that classically describes a black hole. Here we have presented the case when the back-reacted spacetime is a completely regular and geodesically complete causal diamond,
in which the classical horizon is replaced by a wormhole throat.
Second, the particles in the Boulware quantum state may carry the important  information, that is missing when only the particles in the Hartle-Hawking quantum state are considered. 
The ultimate release of the Boulware particles helps to reconstruct the complete information present in the system.


\vspace{0.2 cm}


\noindent{\bf Acknowledgements.}  The work of DS is supported by the DST-FIST grant number SR/FST/PSI-225/2016 and SERB MATRICS grant MTR/2021/000168.


\end{document}